\documentclass[aps,floats,superscriptaddress,showpacs]{revtex4}
\usepackage{graphicx}
\usepackage{overpic}
\usepackage{subfig}
\usepackage{hyperref}

\draft
\usepackage{amsfonts,amssymb}
\usepackage{dsfont}

\newcommand{\bastar}{\begin{eqnarray*}}
\newcommand{\eastar}{\end{eqnarray*}}
\newskip\humongous \humongous=0pt plus 1000pt minus 1000pt

\relax
\newcommand{\be}{\begin{equation}}
\newcommand{\ee}{\end{equation}}
\newcommand{\bea}{\begin{eqnarray}}
\newcommand{\eea}{\end{eqnarray}}

\newcommand{\dfrac}{\displaystyle\frac}

\begin{document}
\title{Optical Simulation of Yang-Baxter Equation}
\bigskip
\author{Shuang-Wei Hu}
\affiliation{Theoretical Physics Division, Chern Institute of
Mathematics, Nankai University, Tianjin 300071, P.R.China}
\author{Ming-Guang Hu}
\affiliation{Theoretical Physics Division, Chern Institute of
Mathematics, Nankai University, Tianjin 300071, P.R.China}
\author{Kang Xue}
\affiliation{Department of Physics, Northeast Normal University,
Changchun, Jilin 130024, P.R.China}
\author{Mo-Lin Ge}
\email{geml@nankai.edu.cn} \affiliation{Theoretical Physics
Division, Chern Institute of Mathematics, Nankai University, Tianjin
300071, P.R.China}

\begin{abstract}
In this paper, several proposals of optically simulating Yang-Baxter
equations have been presented. Motivated by the recent development
of anyon theory, we apply Temperley-Lieb algebra as a bridge to
recast four-dimentional Yang-Baxter equation into its
two-dimensional counterpart. In accordance with both
representations, we find the corresponding linear-optical
simulations, based on the highly efficient optical elements. Both
the freedom degrees of photon polarization and location are utilized
as the qubit basis, in which the unitary Yang-Baxter matrices are
decomposed into combination of
actions of basic optical elements. %The total proposals are fully
%presented to directly simulate Yang-Baxter equations.
\end{abstract}
\vspace{0.3cm} \pacs{05.50.+q, 42.50.Ex, 05.30.Pr}
% pacs: ybe, quantum optics, fractional statistics
%\bigskip
\maketitle

\section{Introduction}

Yang-Baxter equation (YBE) originated in solving the one-dimentional
$\delta$-interacting models \cite{yang} and the statistical models
on lattices \cite{baxter,jimbo}. The importance of YBE was further
realized as a starting point for the quantum inverse scattering
method \cite{faddeev}. It is well-known that YBE plays important
role in solving the integrable models in quantum field theory and
exactly solved models in statistical mechanics (\cite{jimbo} and
references therein). In quantum field theory, YBE describes the
scattering of particles in (1+1) dimensions. The essence of YBE is
to factorize the scattering of three particles into successive
two-body scattering processes. YBE also plays an important role in
completely integrable statistical models, whose solutions can be
found by means of the nested Bethe Ansatz \cite{mattis}.

Due to its importance, YBE deserves thus to be tested
experimentally. Measuring the spectrum of spin chain, which is at
the root of YBE, one can learn the structure of spinon
% (an even number of spin-$1/2$
%objects, \cite{spinon})
and thus check the factorization of YBE. For instance, Heisenberg
spin-1/2 chain model has been probed experimentally through neutron
scattering experiments and the spectrum coincides with the
calculation based on YBE \cite{tennant}.
%However, the directly experimental check of YBE has been absent so
%far due to the complexity of practical physics processes.
However, YBE only provides sufficient condition for the prediction
of spectrum. So these experiments should be viewed as indirect check
of YBE. The direct verification is still an open question.

%Consider a lattice model defined on $L$ sites, each of which has a
%Hilbert space $\mathcal{V}$.

\begin{figure}
{\includegraphics[width=4.5in]{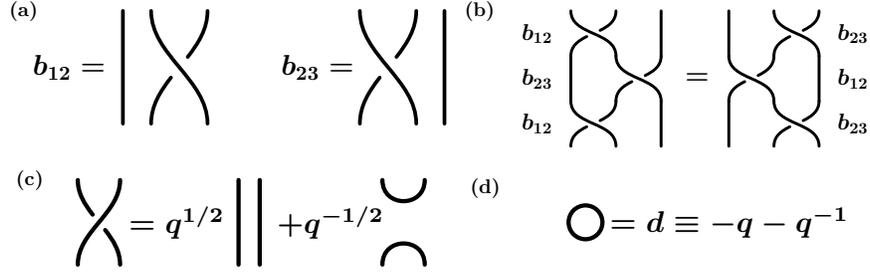}}
%\centering \subfigure{
% \label{fig:subfig:a}
%\includegraphics[width=2in]{b1b2.eps}}
%\hspace{0.7in}
% \subfigure{ \label{fig:subfig:b}
%\includegraphics[width=2in]{aba.eps}}
%\\
%\centering \subfigure{
% \label{fig:subfig:c}
%\includegraphics[width=1.8in]{Kauffman.eps}}
%\hspace{0.8in}
% \subfigure{ \label{fig:subfig:d}
%\includegraphics[width=1.4in]{d.eps}}
\caption{(a) Diagrammatical interpretation of braid operators The
labeled particle worldlines, orienting from bottom to ceiling, are
unaffected by smooth deformations in which the lines do not
intersect. Each crossing means a scattering of two particles,
including permutation process as the special case. (b) The braid
relation (\ref{braid4}) $b_{12}b_{23}b_{12}=b_{23}b_{12}b_{23}$. (c)
Skein relation. (d) The unknotted loop. By skein relation we can
eliminate all the crossings and get a linear combinations of the
Kauffman brackets for various disjoint unions of unknotted loops.}
\label{braidf}
\end{figure}

In order to keep the paper self-contained, we first explain the
basic formula of YBE. The Yang-Baxter matrix $\breve{R}$ is a
$N^2\times N^2$ matrix acting on the tensor product space
$\mathcal{V}\otimes \mathcal{V}$, where $N$ is the dimension of
$\mathcal{V}$. Such a matrix $\breve{R}$ satisfies the YBE
\begin{eqnarray}
\breve{R}_{12}(u)\breve{R}_{23}(\frac{u+v}{1+ \beta^2
uv})\breve{R}_{12}(v)= \breve{R}_{23}(v)\breve{R}_{12}(\frac{u+v}{1+
\beta^2 uv})\breve{R}_{23}(u). \label{315}
\end{eqnarray}
where $\breve{R}_{12}=\breve{R}\otimes \mathds{1}$,
$\breve{R}_{23}=\mathds{1}\otimes \breve{R}$, $u$ and $v$ are
spectral parameters, $\beta^{-1}=i c$ ($c$ is light velocity). Take
the two spin-1/2 particles as an example. Acting on such a system,
$\breve{R}$ is a $2^2\times 2^2$ matrix whose matrix elements are
$\breve{R}_{ab,cd}$, $a,b,c,d=\uparrow$ (spin up), $\downarrow$
(spin down). If spin is conserved, some matrix elements may vanish.
The physical meaning of $\breve{R}(u)$ is two-particle scattering
matrix depending on the relative rapidity $\tanh^{-1} (\beta u)$.
When we change the spectral parameters as $\beta
u=(1-x)/(1+x),~\beta v=(1-y)/(1+y)$ and $\beta(u+v)/(1+\beta^2
uv)=(1-xy)/(1+xy)$, we obtain another ordinary form of YBE
\bea \breve{R}_{12}(x)\breve{R}_{23}(xy)\breve{R}_{12}(y)
=\breve{R}_{23}(y)\breve{R}_{12}(xy)\breve{R}_{23}(x),
\label{1.2}\eea
i.e. the spectral parameter in the middle $\breve{R}(xy)$-matrix
being the product of the neighborhoods' spectral parameters. The
asymptotic limit $ \breve{R}(x\rightarrow 0)=b$ satisfies the braid
relation
\bea b_{12} b_{23} b_{12} =  b_{23}b_{12} b_{23}. \label{braid4}
\eea
This relation is diagrammatically represented in Fig.
(\ref{braidf}). For a given matrix $b$ satisfying (\ref{braid4}) we
can retrieve the corresponding $\breve{R}(x)$-matrix via the
procedure of Baxterization or Yang-Baxterization \cite{yb}. This
procedure depends on the numbers of independent eigenvalues of
matrix $b$. In particular, when a braid matrix $b$ has two
independent eigenvalues $\lambda_1$ and $\lambda_2$, the
corresponding $\breve{R}(x)$-matrix obtained via Yang-Baxterization
takes the form
\bea \breve{R}(x)= \rho(x)(b + x\lambda_1 \lambda_2 b^{-1}),
\label{RB}\eea
where $\rho(x)$ is normalization factor. For statistical models on
lattice, the elements of $\breve{R}(x)$ should be positive-definite,
since they are related to the Boltzmann weights. However, as we will
see below, there is no such a restriction for the application to
quantum entangled states.

%%
%\begin{figure}
%\begin{minipage}[c]{0.5\linewidth}
%\centering\includegraphics[width=1.9in]{b1b2.eps}
%\end{minipage}%
%\begin{minipage}[c]{0.5\linewidth}
%\centering\includegraphics[width=1.9in]{aba.eps}
%\end{minipage}
%\protect\\
%\begin{minipage}[l]{0.5\linewidth}
%\includegraphics[width=1.8in]{Kauffman.eps}
%\end{minipage}%
%\begin{minipage}[l]{0.5\linewidth}
%\includegraphics[width=1.3in]{d.eps}
%\end{minipage}
%\caption{Up: Diagrammatical interpretation of braid operators and
%braid relation (\ref{braid4}). The labeled particle worldlines,
%orienting from bottom to ceiling, are unaffected by smooth
%deformations in which the lines do not intersect. Each crossing
%means a scattering of two particles, including permutation process
%as the special case. Down: Skein relation and an unknotted loop. By
%skein relation we can eliminate all the crossings and get a linear
%combinations of the Kauffman brackets for various disjoint unions of
%unknotted loops.} \label{braidf}
%\end{figure}

In the recent years there is a new development to connect the braid
matrix, as well as YBE, with the entangled state
\cite{kau1,wang,kitaev,freedman,zkg,zg,ckg,zwg}. The basic idea
comes from the Bell states having the maximal entanglement degree.
For a two-qubit system, Bell states are defined by
\bea
&&|\Phi^{\pm}\rangle=\dfrac{1}{\sqrt{2}}(|\uparrow\uparrow\rangle\pm|\downarrow\downarrow\rangle),\nonumber\\
&&|\Psi^{\pm}\rangle=\dfrac{1}{\sqrt{2}}(|\uparrow\downarrow\rangle\pm|\downarrow\uparrow\rangle).
\eea
 The Bell states are connected to the natural basis
$|\Psi_0\rangle=\big{(}|\uparrow\uparrow\rangle,|\uparrow\downarrow\rangle,|\downarrow\uparrow\rangle,|\downarrow\downarrow\rangle\big{)}$
by a unitary transformation matrix $W$,
\bea
&&\big{(}|\Phi^{-}\rangle,|\Psi^{+}\rangle,|\Psi^{-}\rangle,|\Phi^{+}\rangle
\big{)}=W\big{(}|\uparrow\uparrow\rangle,|\uparrow\downarrow\rangle,|\downarrow\uparrow\rangle,|\downarrow\downarrow\rangle\big{)}
,\nonumber
\\
&& W=\dfrac{1}{\sqrt{2}}\left(
\begin{array}{cccc}
1 & 0 & 0 & 1 \\
0 & 1 & -1& 0 \\
0 & 1 & 1 & 0 \\
-1& 0 & 0 & 1
\end{array}
\right). \eea
(We refer the reader to \cite{ckg} for more details about the short
notation in the first line equation.) L. H. Kauffman \emph{et al}
have shown that the matrix $W$ is nothing but a braid matrix
satisfying (\ref{braid4}) by recognizing $\mathcal{V}$ as two
dimensional complex vector space to hold a single qubit of
information \cite{kau1}. Further, it was found that $W$ can be
extended to matrix $b$ such that \cite{ckg}
\bea b(q)=\dfrac{1}{\sqrt{2}}\left(
\begin{array}{cccc}
1 & 0 & 0 & q \\
0 & 1 & -1& 0 \\
0 & 1 & 1 & 0 \\
-q^{-1}& 0 & 0 & 1
\end{array}
\right)=\dfrac{1}{\sqrt{2}}(\mathds{1}+M),~M^2=-\mathds{1},~q=e^{i\phi},\label{bri}
\eea
where real parameter $\phi$ is time-dependent flux. Yang-Baxterizing
this braid matrix $b$, one can define a new state with arbitrary
entanglement degree, as follows
\bea |\Psi(x,q)\rangle =
\breve{R}(x,q)|\Psi_0\rangle,~\breve{R}(x,q) =
\dfrac{1}{\sqrt{1+x^2}}\big{(}b(q)+xb(q)^{-1}\big{)},\label{enta}
\eea
where $\breve{R}(x,q)$-matrix satisfies YBE (\ref{1.2}) and
meanwhile determines the evolution of the initial state
$|\Psi_0\rangle$ to $|\Psi(x,q)\rangle$ given the time-dependent
$q=q(t)$. In terms of the new variable $\cos \Theta =
(1+x)/(\sqrt{2(1+x^2)})$, $\breve{R}(x,q)$ can be recast to
\bea \breve{R}(\Theta,\phi)=\left(
\begin{array}{cccc}
\cos\Theta & 0 & 0 & e^{-i\phi}\sin\Theta \\
0 & \cos\Theta & -\sin\Theta & 0 \\
0 & \sin\Theta & \cos\Theta & 0 \\
-e^{i\phi}\sin\Theta & 0 & 0 & \cos\Theta
\end{array}
\right). \label{bi}\eea
It is interesting to observe that $\breve{R}(\Theta,\phi)$ satisfies
to the relation
\bea \breve{R}(\Theta,\phi)=\cos \Theta \mathds{1} + \sin \Theta
M,~M^2=-\mathds{1}, \eea
i.e. as the extension of the Euler formula.
When $\Theta=\pi/4$, $\breve{R}(\Theta,\phi)$ reduces to $b$ in
(\ref{bri}), i.e. yielding the maximum of entangled states. When
$\Theta$ takes other values the state $|\Psi(\Theta,\phi)\rangle =
\breve{R}(\Theta,\phi)|\Psi_0\rangle$ processes a continuous
entanglement degree determined by $\Theta$ which is usually less
than the maximum \cite{ckg}. Suppose only the flux $\phi$ depends on
$t$ and under the adiabatic approximation, then we can obtain the
Berry phase related to YBE \cite{ckg}.

The theory sounds reasonable, but why we prefer to choose a
Yang-Baxterized matrix $\breve{R}(\Theta,\phi)$ as the unitary
evolution is just an assumption. Especially, one may doubt the
necessity of introducing the third particle to describe two-particle
entanglement. We have to present a practical scheme to test the YBE
in the framework of quantum information. Fortunately, there have
been popular optical operations for the quantum gates \cite{kok}
that are available to experimentally test the validity of such a
YBE. The motivation of this paper is to propose a linear optical
simulation of YBE based on the highly efficient optical elements.

Both the braid matrix $B$ in (\ref{bri}) and the
$\breve{R}(x)$-matrix in (\ref{bi}) act on the tensor product space
$\mathcal{V}\otimes \mathcal{V}$ and thus have the 4-dimensional
(4D) representation. The entangled state in (\ref{enta}) requires
that the optical simulation of the corresponding $\breve{R}$-matrix
involves the universal entangled gate, i.e. CNOT gate. In principle
CNOT gates make use of measurement-induced nonlinearity and are
still of low efficiency by means of optics \cite{kok,brien}. The
situation becomes worse when several sequent CNOT gates are
involved. We have to find alternative ways to avoid this
difficulty. %Fortunately the 2-dimensional (2D) braid behavior under
%the exchange of anyons \cite{wil} has been investigated based on the
%fractional quantum Hall effect (FQHE) \cite{freedman}. Motivated by
%this interesting application of braid relation in anyon theory, we
%will nest Temperley-Lieb algebra into 4D YBE and reduce it to 2D
%YBE.
Fortunately, as we will see in Sec.\ref{sec2}, the 4D Yang-Baxter
matrices have two-dimensional (2D) counterparts which are unitary
and have much simpler realization by means of linear optics.

The paper is organized as the following. In Sec.\ref{sec2}, we first
prove the equivalence between 4D braid matrix and 2D braid matrix,
then find the Yang-Baxterization procedure for 2D braid matrix.
Based on this theoretical assertion the optical test of 2D YBE will
be presented in Sec.\ref{sec3}. A direct test of 4D YBE is shown in
Sec.\ref{sec4}. The conclusion is made in the Sec.\ref{con}. The
relationship between the basic for 2D braid matrices and 4D ones
will be given in Appendix \ref{appena}.

\section{Two Types of YBE's} \label{sec2}

%the 2-dimensional
%braid behavior under the exchange of anyons \cite{wil} has been
%investigated based on the fractional Hall effect \cite{freedman}.
%Motivated by this interesting application of braid relation in anyon
%theory, we nest Temperley-Lieb algebra into 4D YBE and reduce it to
%2D YBE.

In the topological quantum computation theory, the 2D braid behavior
under the exchange of anyons \cite{wil} has been investigated based
on the $\nu=5/2$ fractional quantum Hall effect (FQHE)
\cite{freedman}. Motivated by this interesting application of braid
relation in anyon theory, we will nest Temperley-Lieb algebra
\cite{tl} into 4D YBE and reduce it to 2D YBE. Here we briefly
present such an equivalence between these types of YBE's.

Let us first recall the braid behavior in $\nu=5/2$ FQHE.
Quasiparticles in such a system are often called Ising anyons or
SU(2)$_2$ states, which satisfy non-Abelian fractional statistics.
There are three types of anyons, which can be called 0,
$\frac{1}{2}$, 1. When two anyons become close together while other
anyons are much farther away, these two anyons can be treated as a
single particle whose quantum numbers are obtained by combining the
original quantum numbers. For SU(2)$_2$ states, such a formation of
new anyons obeys the following fusion rules
\bea &&\dfrac{1}{2}\times\dfrac{1}{2} = 0+1,~\dfrac{1}{2}\times 1 =
\dfrac{1}{2},~1\times 1 =0, \nonumber \\
&&0\times 0 =0,~0\times \dfrac{1}{2}= \dfrac{1}{2},~0\times 1 =1.
\label{fusion} \eea
(These fusion rules are analogous to the decomposition rules for
tensor products of irreducible SU(2) representations, but have an
important difference that 1 is the maximum spin.) Note that there
are two different fusion channels for two $\frac{1}{2}$ anyons. As a
result, when four $\frac{1}{2}$ anyons fuse together to give 0,
there is a two-dimensional space of such states. This can be done by
dividing the four $\frac{1}{2}$ anyons into two pairs. Both pairs
either fuse to 0 or to 1 then fuse the resulting anyons together to
form 0.
%If we
%divided the four $\frac{1}{2}$'s into two pairs, by grouping
%particles 1, 2 and 3, 4, then a basis for the two-dimensional space
%is given by the state in which 1, 3 fuse to 1 or 1, 3 fuse to ¦× (2,
%4 must fuse to the same particle type as 1, 3 do in order that all
%four particles fuse to 1). We can call these states     1 and   ¦×;
%they are a basis for the four-quasiparticle Hilbert space with total
%topological charge 1. (Similarly, if they all fused to give ¦×,
%there would be another two-dimensional degenerate space; one basis
%is given by the state in which the first pair fuses to 1 while the
%second fuses to ¦× and the state in which the opposite occurs.)
%
 %Assume
%that operator $A$ makes crossing between the first and second
%quasiparticles (anyons), whereas $B$ makes crossing of the second
%and third quasiparticles. The anyon fusion rules
%($1\otimes\tau=\tau\otimes 1=\tau,~\tau\otimes \tau=1\oplus\tau$)
%define a 2-level systems, $|0\rangle=1,~|1\rangle=\tau$.
%%
%The logic qubit, however, is encoded in doubled Hilbert space formed
%by four quasiparticles.
The orthogonal basis states read \cite{freedman}
\bea
       && {\includegraphics[width=2.3in]{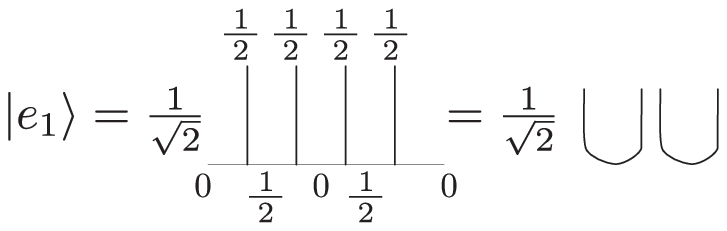}},\nonumber \\
       && {\includegraphics[width=2.8in]{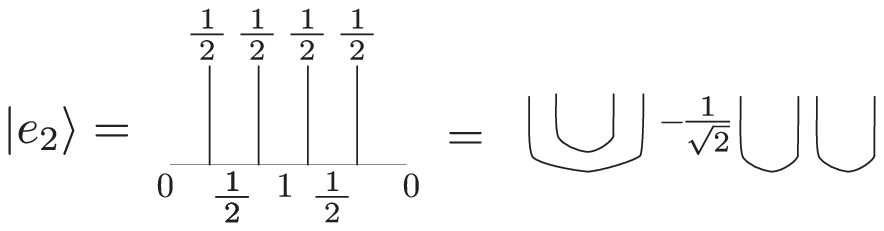}}.
        \label{3.2}
\eea
In the middle fusion chains (called conformal block), the internal
edges are subject to the fusion rules at each trivalent vertex. In
such conformal block basis, exchanging anyons is identified as
braiding in Fig. \ref{braidf}. From the conformal basis to the
Kauffman graph in the right-hand sides, Jones-Wenzl projector
operators have been applied, i.e.
\bea
        {\includegraphics[width=2in]{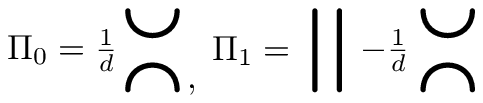}}~,~~
%        {\includegraphics[width=1.2in]{f12.eps}}~,
        \label{3.2}
\eea
where $d=\sqrt{2}$ in present case. By means of skein relation in
Fig. \ref{braidf}, one can introduce the braid operators $A$ and $B$
which have nontrivial action on $|e_1\rangle$ and $|e_2\rangle$
\begin{equation}
    \includegraphics[width=3.2in]{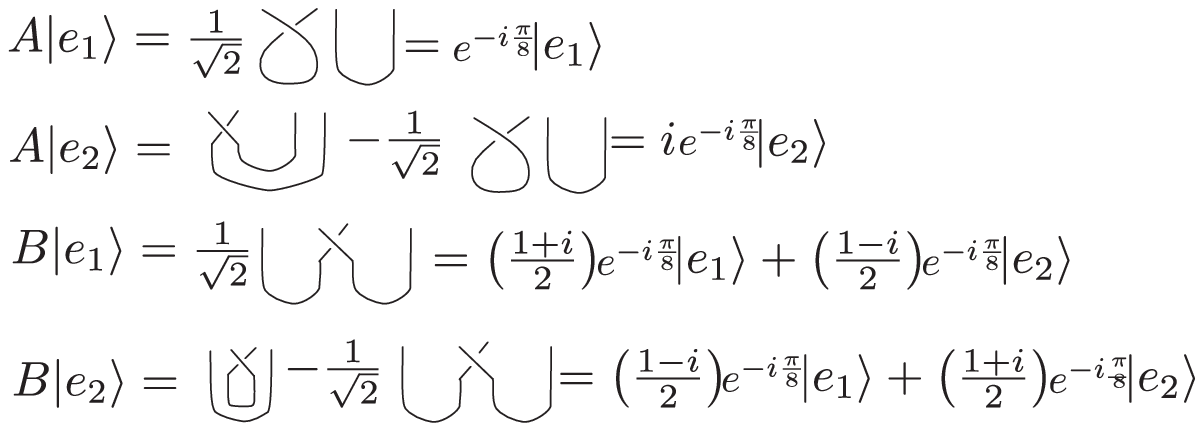}.
    \label{3.3}
\end{equation}
Thus their matrix representations in the basis $\left( |e_1\rangle,~
|e_2\rangle \right)$ are given by
\begin{equation}
    A=e^{-i\frac{\pi}{8}}\left(
    \begin{array}{cc}
        1&0\\
        0&i
    \end{array}
    \right),
    ~
    B=\frac{e^{-i\frac{\pi}{8}}}{2}\left(
    \begin{array}{cc}
        1+i&1-i\\
        1-i&1+i
    \end{array}
    \right),
    \label{3.4}
\end{equation}
and they satisfy 2-dimensional braid relation
\begin{equation}
    ABA=BAB.
    \label{3.5}
\end{equation}
We emphasize that (\ref{3.5}) act on the basis $\left( |e_1\rangle,~
|e_2\rangle \right)$. It is worthy of noting that the ``crossing''
in (\ref{3.3}) means the usual $4\times4$ braid matrix, satisfying
(\ref{braid4}).  These braid matrices $A$ and $B$ are unitary and
have a natural realization by
linear optics, as we will see in Sec.\ref{sec3}. %Up to a phase, matrix $B$ is the action of quarter
%wave plate (QWP) which is oriented with its fast axis at $\pi/4$ to
%the vertical direction.

In order to generalize the above procedure from braid relation to
YBE, we nest Temperley-Lieb algebra \cite{tl} into 4D YBE and
surprisingly reduce it to 2D YBE. Detailed calculations will be
given later (see Appendix \ref{appena}). %By re-parameterization,
%$u=-i\tan r,~v=-i\tan s$, the 4D YBE (\ref{315}) can be rewritten as
%%
%\begin{eqnarray}
%\breve{R}_{12}(u)\breve{R}_{23}(\frac{u+v}{1+ uv})\breve{R}_{12}(v)=
%\breve{R}_{23}(v)\breve{R}_{12}(\frac{u+v}{1+ uv})\breve{R}_{23}(u).
%\label{315}
%\end{eqnarray}
%
Briefly, acting on the subspace spanned by $|e_1\rangle$ and
$|e_2\rangle$, the 4D YBE (\ref{315}) will reduce to the
corresponding 2D YBE
\bea &&A(u)B(\frac{u+v}{1+ \beta^2 u v})A(v)=B(v)A(\frac{u+v}{1+
\beta^2 u
v})B(u), \nonumber \\ %\label{3.20}
%\end{equation}
%%
%And $A(u)$ and $B(u)$ take the form as follows,
%\begin{eqnarray}
&&A(u)=\rho(u)\left( \begin{array}{cc} \dfrac{1+\beta^2
u^2+2i\epsilon \beta u}{1+ \beta^2 u^2-2i\epsilon \beta u}&0\\0&1
\end{array} \right),
\nonumber \\
&&B(u)=\frac{\rho(u)}{1+ \beta^2 u^2- 2i\epsilon \beta u}\left(
\begin{array}{cc}1+ \beta^2 u^2 & 2i\epsilon \beta
u\\2i\epsilon \beta u & 1 + \beta^2 u^2 \end{array} \right),
\label{3.19}
\end{eqnarray}
where $\rho(u)$ is normalization factor and $\epsilon=\pm1$. Since
these matrices $A(u)$ and $B(u)$ are unitary, it is obviously easier
to optically simulate this 2D equation than the previous 4D edition.
For the convenience of experimental check, we further introduce the
transformation
\bea \dfrac{1+ \beta^2u^2+2i\epsilon \beta u}{1+ \beta^2
u^2-2i\epsilon \beta u}\equiv e^{-2i\theta},~\rho(u)\equiv
e^{i\theta},\label{equiv} \eea
 then we
obtain the following form of SU(2) matrices
\bea A(u) &=& \left(
\begin{array}{cc}
e^{-i\theta} & 0 \\
0 & e^{i\theta}
\end{array}
\right)\equiv A(\theta), \nonumber \\
B(u)&=&\left(
\begin{array}{cc}
\cos \theta & -i\sin\theta \\
-i\sin\theta & \cos\theta
\end{array}
\right)\equiv B(\theta).\label{ori}\eea
In terms of this new parameters, the solution of
$\breve{R}(\theta)$-matrix takes the form
\bea \breve{R}(\theta,\phi)=\left(
\begin{array}{cccc}
\cos\theta & 0 & 0 & e^{-i\phi}\sin\theta \\
0 & \cos\theta & -\sin\theta & 0 \\
0 & \sin\theta & \cos\theta & 0 \\
-e^{i\phi}\sin\theta & 0 & 0 & \cos\theta
\end{array}
\right). \label{R}\eea
Though it takes the similar form with (\ref{bi}), the parameter
$\theta$ has different meaning from the parameter $\Theta$. One may
wonder where is the missing parameter $\phi$ in 2D YBE. It actually
survives in the basis $|e_1\rangle$ and $|e_2\rangle$ which 2D YBE
should act on (see the explicit form of the basis in Appendix
\ref{appena}).
\begin{figure}[ht]
\includegraphics[width=0.5\columnwidth]{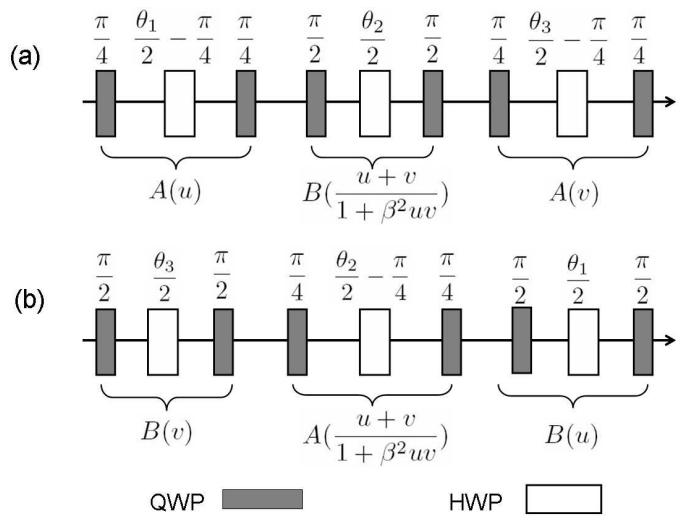}
\caption{Schematic setup for simulating either side of 2D YBE
(\ref{3.19}) by means of polarization qubit. The angle parameters
$\theta_i$ are determined by (\ref{para}) while the relations
between $\theta_i$ and $u,~v$ are determined by (\ref{para0}). }
\label{u3}
\end{figure}

These two unitary matrices can be optically realized with the aid of
the "polarization qubit" or "location qubit" of a single photon. In
the following we present two experimental setups to simulate YBE by
means of these two ways.

\section{Optical Simulation of 2D YBE} \label{sec3}

The single-photon representation of qubits plays an important role
in linear optical computation \cite{pra,cerf,kuang}. The key is that
a single photon can be utilized to act both as polarization qubit
and as location qubit. For the former one encodes the qubit in the
photon's polarization with the corresponding transformations
simulated by wave plates, such as half-wave plates (HWPs) and
quarter-wave plates (QWPs). For the latter one encodes the qubit in
the single-photon paths with the corresponding transformations
simulated by beam splitters (BSs), phase shifters (PSs), and
mirrors. As a result, universal unitary gates (1-qubit or 2-qubit)
can be realized by means of either of these two kinds of qubit
transformation, or by combining both \cite{pra,cerf}.

The two unitary matrices in (\ref{ori}) can be optically realized
with the aid of the "polarization qubit" or "location qubit" of a
single photon. In the following we present two experimental setups
to simulate YBE by means of these two ways.

\subsection{Using Polarization Qubit to Simulate 2D YBE}

We first recall the action of a QWP upon the basis states of the
polarization qubit \cite{kuang}
\bea U_Q(\delta)= e^{-i\delta\sigma_2} e^{-i(\pi/4)\sigma_3}
e^{i\delta\sigma_2} = \frac{1}{\sqrt{2}} \left(
\begin{array}{cc}
1-i\cos(2\delta) & -i\sin(2\delta) \\
-i\sin(2\delta) & 1+i\cos(2\delta)
\end{array}
\right), \eea
where $\sigma_i$ are Pauli matrices and $\delta$ is the angle
between the QWP axis and the vertical direction. Then the action of
a HWP upon the basis states of the polarization qubit is given by
\bea U_H(\delta)= U_Q^2(\delta)= -i \left(
\begin{array}{cc}
\cos(2\delta) & \sin(2\delta) \\
\sin(2\delta) & -\cos(2\delta)
\end{array}
\right). \eea

%$\dfrac{\theta_1}{2}-\dfrac{\pi}{4}~~\dfrac{\theta_3}{2}-\dfrac{\pi}{4}~~\dfrac{\theta_2}{2}$

As an analogue with Euler rotation, the sandwich configuration of
one HWP and two QWPs enables one to perform any unitary changes of
the photon¡¯s polarization state \cite{book}. Particularly, for the
case in (\ref{ori}), we obtain
\bea A(\theta) = U_Q(\frac{\pi}{4})
U_H(-\frac{\pi}{4}+\frac{\theta}{2}) U_Q(\frac{\pi}{4}), ~B(\theta)
= U_Q(\frac{\pi}{2}) U_H(\frac{\theta}{2}) U_Q(\frac{\pi}{2}).
\label{exp}\eea
By this decomposition it is straightforward to design the
experimental setup for simulation of 2-dimensional YBE (\ref{3.19}).
As Fig.\ref{u3} shows, a suitable series of QWPs and HWPs with
different direction angles in succession will simulate the left-hand
side (LHS) of YBE while another series will do the right-hand side
(RHS).
In Fig.\ref{u3}, the relation (\ref{equiv}) requires the angle
parameters $\theta_i$ in the LHS satisfy
\bea \dfrac{1-\beta^2u^2+2i\epsilon \beta u}{1-\beta^2
u^2-2i\epsilon \beta u}=e^{-2i\theta_1},
~\dfrac{1-\beta^2(\dfrac{u+v}{1+\beta^2 u v})^2+2i\epsilon \beta
\dfrac{u+v}{1+ \beta^2 u v}}{1- \beta^2(\dfrac{u+v}{1+\beta^2 u
v})^2-2i\epsilon \beta \dfrac{u+v}{1+\beta^2 u v}} =e^{-2i\theta_2},
~\dfrac{1-\beta^2v^2+2i\epsilon \beta v}{1-\beta^2 v^2-2i\epsilon
\beta v}=e^{-2i\theta_3}.\label{para0} \eea
Thus we have
\bea \big{(}e^{-2i\theta_2}+1\big{)}\big{(}i-\sec(\theta_1-\theta_3)
\sin (\theta_1+\theta_3)\big{)}=2i. \label{para} \eea
This also holds for the angle parameters appearing in RHS of YBE
(\ref{3.19}). What we should measure in experiment is to analysis
the actions of transformation of both sides given the input states
with the same angle parameters, for example, by means of quantum
state tomography \cite{tomo}.

%
%for the left-hand side of YBE (\ref{3.20}) it requires that
%%
%\bea &&
%\beta_1=-\frac{\pi}{4}+\frac{\theta_1}{2},~\beta_2=\frac{\theta_2}{2},
%~\beta_3=-\frac{\pi}{4}+\frac{\theta_3}{2}, \nonumber \\
%&\Rightarrow&\sec2(\beta_1-\beta_3) \sin
%2(\beta_1+\beta_3)+i=\dfrac{2i}{e^{-4i\beta_2}+1}. \label{lhs} \eea
%%
%Similarly for the right-hand side
%%
%\bea &&
%\beta_1=\frac{\theta_1}{2},~\beta_2=-\frac{\pi}{4}+\frac{\theta_2}{2},
%~\beta_3=\frac{\theta_3}{2}, \nonumber \\
%&\Rightarrow&  \sec2(\beta_1-\beta_3) \sin
%2(\beta_1+\beta_3)-i=\dfrac{-2i}{-e^{-4i\beta_2}+1}. \label{rhs}
%\eea
%
\begin{figure}[ht]
\includegraphics[width=0.85\columnwidth]{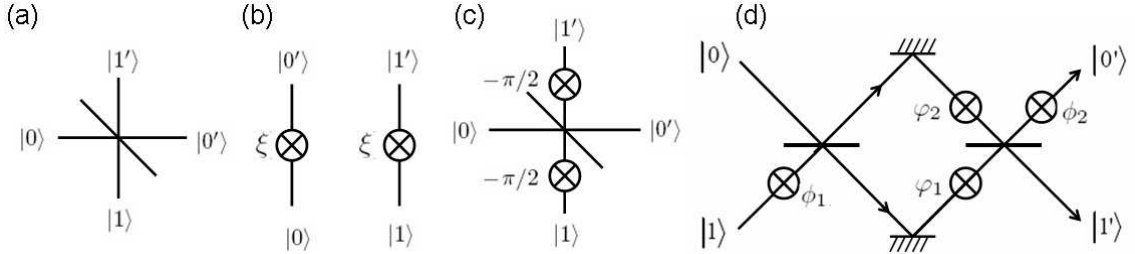}
\caption{Gates acting on location qubit basis: (a) Beam splitter
$U_{BS}$. (b) Phase shifters $U^0_{PS}(\xi)$ (left) and
$U^1_{PS}(\xi)$ (right). (c) Hadamard gate $H$ by a 50:50 BS and two
$-\pi/2$ PS. (d) Mach-Zehnder interferometer $U_{MZ}$ as universal
1-qubit gate. For simplicity, each pair of $-\pi/2$ phase shifters
accompanied with every beam splitter as in (c) is not shown in (d).
} \label{elel}
\end{figure}

\subsection{Using location qubit to simulate 2D YBE}

\begin{figure}[ht]
\includegraphics[width=0.6\columnwidth]{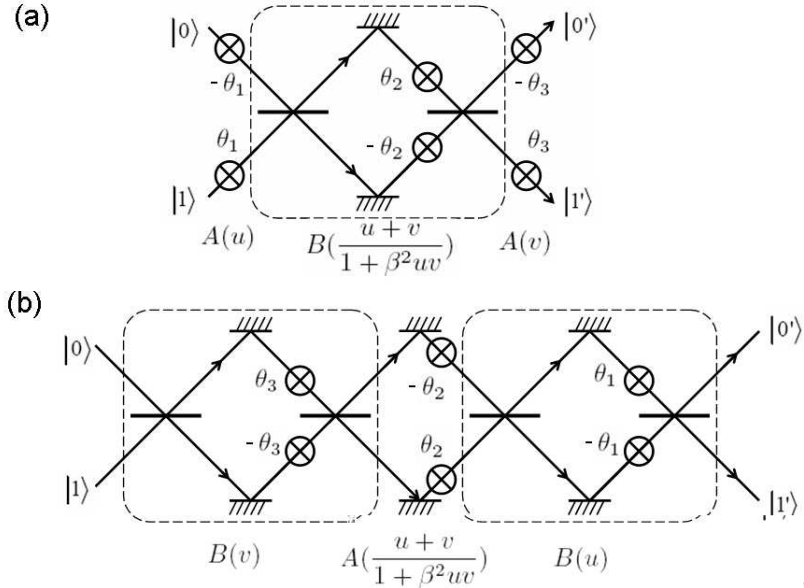}
\caption{Schematic setup for simulating either side of 2D YBE
(\ref{3.19}) by means of location qubit. (a) Simulation of LHS. (b)
Simulation of RHS. Each pair of $-\pi/2$ phase shifters accompanied
with every beam splitter is not shown. The relations of different
parameters are refereed to (\ref{para}).} \label{loc}
\end{figure}

When we take photon location paths as qubit basis, the unitary
transformations can be achieved by means of BSs and PSs. We follows
the notations in \cite{cerf} which are different from those in
\cite{pra}, especially the opposite definitions of location qubit
lead to distinct actions of mirror. First, we list the actions of
several elementary gates on the location qubit basis (see
Fig.\ref{elel})
\bea && U_{BS}=\frac{1}{\sqrt{2}} \left(
\begin{array}{cc}
1 & i \\
i & 1 \end{array} \right),~U_{mirr}=\mathds{1}_2,~
U^0_{PS}(\xi)=\left(
\begin{array}{cc}
e^{i\xi} & 0 \\
0 & 1 \end{array} \right), ~ U^1_{PS}(\xi)=\left(
\begin{array}{cc}
1 & 0 \\
0 & e^{i\xi} \end{array} \right), \nonumber \\
&& H =U^1_{PS}(-\frac{\pi}{2}) U_{BS} U^1_{PS}(-\frac{\pi}{2}) =
\frac{1}{\sqrt{2}} \left(
\begin{array}{cc}
1 & 1 \\
1 & -1 \end{array} \right) .\eea
%
%\begin{figure}[ht]
%\includegraphics[width=0.34\columnwidth]{mz.eps}
%\caption{Mach-Zehnder interferometer: the two beam splitters have
%the ratio of transmission over reflection coefficients as 50 : 50;
%the phase shifters have the phase shifts $\phi_1$, $\phi_2$,
%$\varphi_1$ and $\varphi_2$ respectively; each pair of $-\pi/2$
%phase shifters accompanied with every beam splitter is not shown.}
%\label{mz}
%\end{figure}
Based on these gates, a Mach-Zehnder interferometer
(Fig.\ref{elel}(d)) can realize arbitrary $U(2)$ group element
\cite{cerf,pra}. Note that in Fig.\ref{elel}(d) each pair of
$-\pi/2$ phase shifters accompanied with every beam splitter is not
shown for simplicity. We hold this convention hereafter, so each BS
should be taken as Hadamard gate. The unitary action of Mach-Zehnder
interferometer is given by
\bea
U_{MZ}=U^1_{PS}(\phi_1)HU_{mirr}U^1_{PS}(\varphi_2)U^0_{PS}(\varphi_1)HU^0_{PS}(\phi_2)
=e^{i\frac{\Phi}{2}} \left(
\begin{array}{cc}
e^{-i\frac{\phi_1-\phi_2}{2}}\cos\frac{\lambda}{2} & ie^{i\frac{\phi_1+\phi_2}{2}}\sin\frac{\lambda}{2} \\
ie^{-i\frac{\phi_1+\phi_2}{2}}\sin\frac{\lambda}{2} &
e^{i\frac{\phi_1-\phi_2}{2}}\cos\frac{\lambda}{2}
\end{array}
\right),\eea
where the total phase $\Phi=\phi_1+\phi_2+\varphi_1+\varphi_2$ and
phase difference $\lambda=\varphi_2-\varphi_1$.
Through Mach-Zehnder interferometer, we can perform the action of
operators $A$ and $B$ in the (\ref{exp}) with angles correspondences
as
\bea &&
A(\theta)=U_{MZ}(\varphi_2=\varphi_1=0,~\phi_1=-\phi_2=\theta)=U^0_{PS}(-\theta)U^1_{PS}(\theta),
\nonumber \\&&
B(\theta)=U_{MZ}(\varphi_2=-\varphi_1=\theta,~\phi_1=\phi_2=0).
\label{parainloc} \eea
Then we come to the whole optical setup to simulate both sides of 2D
YBE, (\ref{3.19}), as shown in Fig.\ref{loc}. The angle parameters
obey the same relation in (\ref{para}).

\section{Optical Simulation of four Dimensional YBE}
\label{sec4}

The YBE in four dimension representation (\ref{315}) is equivalent
to
\bea (\breve{R}(\theta_1) \otimes \mathds{1}_2)\cdot(\mathds{1}_2
\otimes \breve{R}(\theta_2))\cdot (\breve{R}(\theta_3) \otimes
\mathds{1}_2) = (\mathds{1}_2\otimes \breve{R}(\theta_3))\cdot
(\breve{R}(\theta_2) \otimes \mathds{1}_2)\cdot(\mathds{1}_2 \otimes
\breve{R}(\theta_1)), \eea
where $\theta_i$ satisfy the relation (\ref{para}) and
$\breve{R}(\theta)$ takes the form in (\ref{R}).
%
%\bea \breve{R}(\theta) = \left(
%\begin{array}{cccc}
%\cos \theta & 0 & 0 & e^{-i\phi} \sin \theta \\
%0 & \cos \theta & -\sin \theta & 0 \\
%0 & \sin \theta & \cos \theta & 0 \\
%- e^{i\phi} \sin \theta & 0 & 0 & \cos \theta
%\end{array}
%\right).\eea
%
The $\breve{R}(\theta)$-matrix can be decomposed into the
combination of elementary gates \cite{dec}. The case $\theta=0$ is
trivial. When $\theta=\pi/4$ or $3\pi/4$ (we restrict
$\theta\in(0,2\pi)$), $\breve{R}(\theta)$ reduces to the braid
matrix (\ref{bri}) and thus equivalent to one CNOT gate, as L. H.
Kauffman \emph{et al} first pointed out \cite{kau1}. When $\theta$
takes other values, it can decomposed as follows
\bea &&\breve{R}(\theta)=(V_1 \otimes V_2)\cdot \textmd{CNOT2} \cdot
(V_3 \otimes V_4) \cdot \textmd{CNOT2} \cdot (V_5 \otimes V_6),
\label{dec} \eea
where $V_i \in U(2)$ and CNOT2 gate is given by
\bea \textmd{CNOT2} = \left(
\begin{array}{cccc}
1 & 0 & 0 & 0 \\
0 & 0 & 0 & 1 \\
0 & 0 & 1 & 0 \\
0 & 1 & 0 & 0
\end{array}
\right). \eea
Below we focus on the general case, i.e. $\theta\neq 0,~\pi/4$ or
$3\pi/4$.
Recently, S. S. Bullock and his coauthors \cite{bullock} has
developed a criterion for determining the number of CNOT (or
equivalently CNOT2) gates to simulate a given transformation. By
this criterion, one can check that the decomposition (\ref{dec}) is
optimal, i.e. $\breve{R}(\theta)$-matrix admits a quantum circuit
using 2 CNOT gates (see Appendix \ref{appenc}).
\begin{figure}[ht]
\includegraphics[width=0.53\columnwidth]{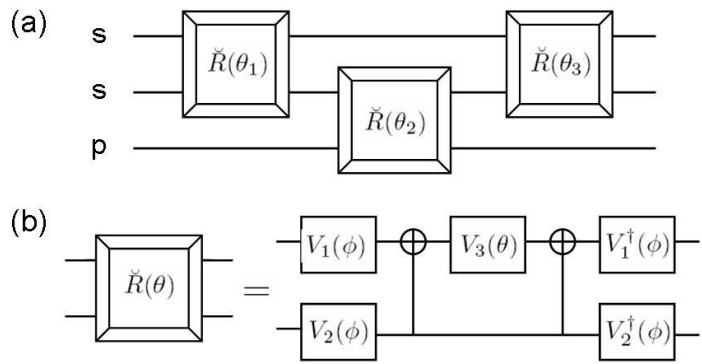}
\caption{(a) A circuit for simulating left-hand side of
4-dimensional YBE (\ref{ybe4}), with s and p denoting location and
polarization qubit channels respectively. (b) the decomposition of
$\breve{R}(\theta)$-matrix, see (\ref{dec}). } \label{ybe4}
\end{figure}

%$V_1(\phi)~~V_2(\phi)~~V_3(\theta)~~V_1^{\dag}(\phi)~~V_2^{\dag}(\phi)$

Detailed calculation (see Appendix \ref{appenc}) further shows $V_i
\in SU(2)$ for the present case. More explicitly, we have
\bea  &&V_1 = \frac{1}{\sqrt{2}} \left(
\begin{array}{cc}
e^{-i\frac{\pi+\phi}{4}} & e^{-i\frac{\pi+\phi}{4}}  \\
-e^{i\frac{\pi+\phi}{4}} & e^{i\frac{\pi+\phi}{4}}
\end{array}
\right),~V_2 = \frac{1}{\sqrt{2}} \left(
\begin{array}{cc}
e^{-i\frac{\phi}{4}} & -e^{-i\frac{\phi}{4}} \\
e^{i\frac{\phi}{4}} & e^{i\frac{\phi}{4}}
\end{array}
\right), ~ V_3 = \left(
\begin{array}{cc}
e^{-i\theta} & 0 \\
0 & e^{i\theta}
\end{array}
\right), \nonumber \\ && V_4 =\mathds{1}_2, ~~V_5=V_1^{\dag},
~~V_6=V_2^{\dag}.\label{ui} \eea
By this explicit decomposition (\ref{dec}), the design of the
circuit to simulate YBE is straightforward, as shown in
Fig.\ref{ybe4}. The difficulty lies in realization of the CNOT
gates. As is described in previous section, photon can carry either
"polarization qubit" or "location qubit". If we only use the former,
CNOT gates are possible for photons in principle using
measurement-induced nonlinearity \cite{knill}. However, currently
they are still low-efficient and experimentally expensive
\cite{kok,brien}. For the present status of linear optics
experiments, it was shown that the success probability of an array
of $n$ CNOT gates can be made to operate with a probability of $ p=
(\frac{1}{3})^{n+1}$ \cite{ralph}. The above decomposition of
$\breve{R}(\theta)$ takes 2 CNOT2 gates. For each side of YBE
(\ref{ybe4}), we have to deal with 6 CNOT gates at the same time.
Thus the success probability is
$p=(\frac{1}{3})^7\simeq4.57\times10^{-4}$, which makes the
practical simulation extremely difficult. This is the reason why we
did map the 4D YBE to 2D YBE.

Photon used as location qubit will help reduce the above difficulty
when we do small-scale quantum calculation. The key lies in the high
efficiency of BSs, PSs and wave plates. In Fig.\ref{ybe4} we have to
deal with three qubit, then two schemes are available: one
polarization qubit channel plus two location qubit channels, or all
location qubit channels. We focus on the former since it save one
optical way and use less number of beam splitters.

\begin{figure}[ht]
\includegraphics[width=0.82\columnwidth]{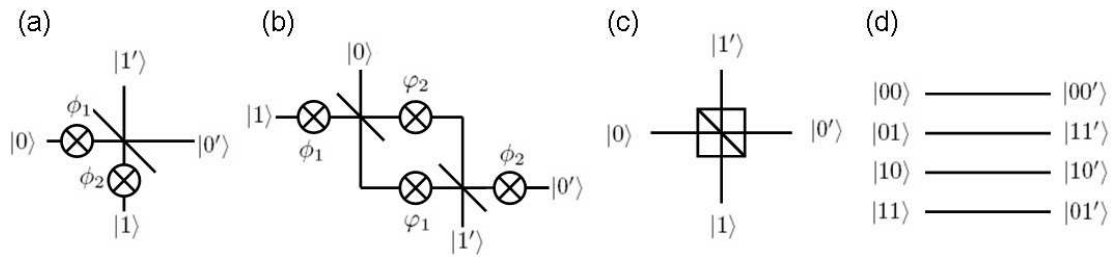}
\caption{(a) The realization of $U_1$ acting on location qubit with
$\phi_1=-\frac{\pi+\phi}{4}$ and $\phi_2=\frac{5\pi+\phi}{4}$. (b)
The realization of $U_2$ acting on location qubit with
$\phi_1=\frac{\pi}{2}$, $\phi_2=\frac{\pi-\phi}{2}$,
$\varphi_1=\frac{\phi-\pi}{4}$ and $\varphi_2=\frac{\phi-3\pi}{4}$.
(c) CNOT2 gate using a polarizing beam splitter with the
polarization and location being the control and target qubit,
respectively. (d) CNOT2 gate between location qubits, which is
achieved by swapping the labels of output paths $|01\rangle$ and
$|11\rangle$.} \label{ele}
\end{figure}

In Fig.\ref{ybe4}(a), three channels are designed to be location
(s), location (s) and polarization (p) qubits from top to floor. For
polarization qubit, using the QWPs and HWPs, the unitary matrices
$V_i$ can be decomposed into
\bea &&V_1= U_Q(\frac{\pi}{4}) U_H(\frac{\phi}{8})
U_Q(\frac{\pi}{2}), ~V_2=  U_Q(-\frac{\pi}{4})
U_H(\frac{\pi-\phi}{8})
U_Q(\frac{\pi}{2}), \nonumber \\
&&V_2^{\dagger}=  U_Q(0) U_H(\frac{5\pi-\phi}{8})
U_Q(\frac{\pi}{4}), ~V_3 = U_Q(\frac{\pi}{4})
U_H(\frac{2\theta-\pi}{4}) U_Q(\frac{\pi}{4}). \label{uii} \eea
For the location qubit, as the previous section shows, a
Mach-Zehnder interferometer can simulate these $V_i$ matrices. We
summarize the results in Fig.\ref{ele}(a) and (b).

%$|00'\rangle~~|01'\rangle~~|10'\rangle~~|11'\rangle$

The rest come to two types of CNOT2 gates: one is between location
qubit and polarization qubit, the other is between two location
qubits. For the former, it can be achieved by a polarizing beam
splitter where the location qubit is flipped or not conditionally on
its state of polarization, as shown in Fig.\ref{ele}(c) \cite{cerf}.
For the latter, the two location qubits in Fig.\ref{ele}(d)
correspond to the first and second number of the binary
representation of the location of a single photon, respectively.
Thus the corresponding CNOT2 is simulated by simply swapping the
labels of path $|10\rangle$ and $|11\rangle$ (see Fig.\ref{ele}(d)).

Gathering all the elementary gates, we finally arrive at the whole
scheme for optically simulating the LHS of 4D YBE, as shown in
Fig.\ref{total}. In order to get the optical setup to simulate the
RHS of 4D YBE, we can apply the formal equivalence between two
hand-sides in the YBE (\ref{315}) by cycling the indices
$1\rightarrow2,~2\rightarrow3,~3\rightarrow1$ and exchanging the
parameters $u\leftrightarrow v$. The equality will be confirmed by
means of tomography given the same input on each sides.

\section{Conclusion} \label{con}

We have presented several proposals to optically simulate
Yang-Baxter equations. According to the development of theoretical
analysis, Yang-Baxter equation in two-dimensional representation and
in four-dimensional representation can be uniformed with the aid of
Temperley-Lieb algebra. In both representations, we have found the
corresponding linear-optical realizations, based on the highly
efficient optical elements, i.e. half-wave plates, quarter-wave
plates, beam splitters, phase shifters, and mirrors. Both the
degrees of freedom of photon polarization and location have been
utilized as the qubit basis. In each kind of basis, the unitary
Yang-Baxter matrices have been decomposed into combination of
actions of basic optical elements. The developed proposals, in
principle, are able to be used to directly check Yang-Baxter
equation.  We remark that the test of 2D YBE is, in fact, to provide
an optical simulation of two-component anyons associated with FQHE.
The anyon behavior is now a hot topic \cite{freedman}, but in our
knowledge there has not been a scheme to test by using an optical
simulation.

        ~~~~~~~~~~~~

   \begin{center}
{\bf Acknowledgements}
\end{center}

  The authors would like to thank Prof. J. L. Chen, Prof. X. Wan, B. X. Xie and M. S. Li for their
  useful discussions. This work was supported in part by NSF of
  China (Grants No. 10575053) and Liu Hui Center of Nankai and
Tianjin Universities.

          ~~~~~~~~~~~~

%   \begin{center}
%{\bf Appendix}
%\end{center}

\appendix

\section{Reduction of 4D YBE} \label{appena}

Analogue with the mapping from 4D braid relation to 2D braid
relation in Sec.\ref{sec2}, we here give the calculation of their
Yang-Baxterized edition. This is equivalent to Yang-Baxterize
(\ref{3.5}).
First we recall the involved YBE
\begin{eqnarray}
\breve{R}_{12}(u)\breve{R}_{23}(\frac{u+v}{1+ \beta^2
uv})\breve{R}_{12}(v)= \breve{R}_{23}(v)\breve{R}_{12}(\frac{u+v}{1+
\beta^2 uv})\breve{R}_{23}(u). \label{3.15}
\end{eqnarray}
%
%where $4\times 4$ matrix $\breve{R}(u)$: $\mathds{C}^2 \otimes
%\mathds{C}^2 \rightarrow \mathds{C}^2 \otimes \mathds{C}^2 $ is an
%invertible linear transformation parametrized by a complex variable
%$u$, $\mathds{C}^2$ is a 2-dimensional complex vector space, and
%$\breve{R}_{12}(u)\equiv \breve{R}(u) \otimes
%\mathds{1}_2,~\breve{R}_{23}(v)\equiv \mathds{1}_2 \otimes
%\breve{R}(v)$.
%
What is interesting is that this four-dimensional YBE (\ref{3.15})
admits the celebrated Temperley-Lieb algebra (TLA). Actually set
\begin{equation}
    \left.
    \begin{array}{ccc}
        \breve{R}_{12}(u)&=&a_1(u)\mathds{1}_6+b_1(u)U_{12}, \\
        \breve{R}_{23}(u)&=&a_2(u)\mathds{1}_6+b_2(u)U_{23},
    \end{array}
    \right.
    \label{3.6}
\end{equation}
and suppose $U$ satisfying TLA \cite{tl}
\begin{equation}
    \left.
    \begin{array}{c}
        U^2=dU,~U_{12}U_{23}U_{12}=U_{12},~U_{23}U_{12}U_{23}=U_{23},
    \end{array}
    \right.
    \label{3.7}
\end{equation}
where $d$ is the loop in Fig.\ref{braidf}, taking the value of
$\sqrt{2}$ in our case. The coefficient functions $a_i(u)$ and
$b_j(u)$ in (\ref{3.6}) are determined by the associated YBE
(\ref{3.15}). Consider one simple but important case:
$a_1(u)=a_2(u)=a(u)$ and $b_1(u)=b_2(u)=b(u)$. We easily get
\begin{eqnarray}
 [a(u)b(v)+b(u)a(v)+d\  b(u)b(v)]a(\frac{u+v}{1+  uv})=
[a(v)a(u)-b(v)a(u)]b(\frac{u+v}{1+  uv}). \label{3.16}
\end{eqnarray}
(\ref{3.16}) has the solution
\bea
 a(u)=\rho(u), ~b(u)=\rho(u)G(u),~G(u)=\frac{4i\epsilon \beta
u}{\sqrt{2}(1+\beta^2 u^2-2i\epsilon \beta  u)} ~(\epsilon=\pm1).
\label{3.17} \eea

On the other hand, in accordance with (\ref{3.2}) and (\ref{3.3}),
the new basis $\left|e_i\right>(i=1,2)$ are introduced from the
definition of the operator $U$
\begin{equation}
    \left.
    \begin{array}{l}
        U_{12}\left|e_1\right>=U_{34}\left|e_1\right>=d\left|e_1\right>,\\
        U_{12}\left|e_2\right>=U_{34}\left|e_2\right>=0,\\
        U_{23}\left|e_1\right>=U_{14}\left|e_1\right>=\dfrac{1}{d}(\left|e_1\right>+\sqrt{d^2-1}\left|e_2\right>),\\
        U_{23}\left|e_2\right>=U_{14}\left|e_2\right>=\dfrac{\sqrt{d^2-1}}{d}(\left|e_1\right>+\sqrt{d^2-1}\left|e_2\right>).
    \end{array}
    \right.
    \label{3.90}
\end{equation}
Thus we have
\begin{equation}
    \left.
    \begin{array}{l}
        \breve{R}_{12}(u)\left|e_1\right>=[a_1(u)+d~b_1(u)]\left|e_1\right>,\\
        \breve{R}_{12}(u)\left|e_2\right>=a_1(u)\left|e_2\right>,\\
        \breve{R}_{23}(u)\left|e_1\right>=[a_2(u)+\dfrac{b_2(u)}{d}]\left|e_1\right>+\dfrac{\sqrt{d^2-1}}{d}b_2(u)\left|e_2\right>,\\
        \breve{R}_{23}(u)\left|e_2\right>=\dfrac{\sqrt{d^2-1}}{d}b_2(u)\left|e_1\right>+[a_2(u)+\dfrac{d^2-1}{d}b_2(u)]\left|e_2\right>.
    \end{array}
    \right.
    \label{3.9}
\end{equation}
So $|e_1\rangle$ and $|e_2\rangle$ span the $\breve{R}$-invariant
subspace. Then it is natural to define the matrix elements
$A(u)_{ij}=\langle e_i|\breve{R}_{12}(u)|e_j\rangle$ and
$B(u)_{ij}=\langle e_i|\breve{R}_{23}(u)|e_j\rangle$.

Combing the above results we obtain the explicit form of $A(u)$ and
$B(u)$,
\begin{eqnarray}
&&A(u)=\rho(u)\left( \begin{array}{cc} \dfrac{1+
\beta^2u^2+2i\epsilon \beta u}{1+ \beta^2 u^2-2i\epsilon \beta
u}&0\\0&1
\end{array} \right),
\nonumber \\
&&B(u)=\frac{\rho(u)}{1+ \beta^2 u^2-2 i \epsilon \beta^2 u}\left(
\begin{array}{cc}1+ \beta^2 u^2 & 2i\epsilon \beta
u\\2 i\epsilon \beta u & 1+ \beta^2 u^2 \end{array} \right),
\label{319}
\end{eqnarray}
and importantly, in accord with the braid relation (\ref{3.5}), they
satisfy the two-dimensional (2D) YBE
\begin{equation}
A(u)B(\frac{u+v}{1+  u v})A(v)=B(v)A(\frac{u+v}{1+  u v})B(u).
\label{320}
\end{equation}
The corresponding unitary matrix $U$ with\ $d=\sqrt{2}$, which
satisfies TLA in (\ref{3.7}), takes the representation
\begin{eqnarray}
U=\frac{1}{\sqrt{2}}\left(
\begin{array}{cccc}1&0&0& iq^{-1}\\0&1&i\epsilon&0\\0&-i\epsilon&1&0\\-iq&0&0&1
\end{array} \right)=\dfrac{1}{\sqrt{2}}(\mathds{1}+iM),~M^2=-\mathds{1},~q=e^{i\phi},~\phi\in \mathds{R}.
\label{3.22}
\end{eqnarray}
Here the important factor $i$ before matrix $M$ distinguishes $U$
from braid operator in (\ref{bri}). In terms of new parameters as in
(\ref{equiv}), the explicit form of $\breve{R}(\theta)$ takes
\bea \breve{R}(\theta) = a(u)+b(u)U =\rho(u)(\mathds{1}_4+G(u)U) =
\left(
\begin{array}{cccc}
\cos \theta & 0 & 0 & e^{-i\phi} \sin \theta \\
0 & \cos \theta & -\sin \theta & 0 \\
0 & \sin \theta & \cos \theta & 0 \\
- e^{i\phi} \sin \theta & 0 & 0 & \cos \theta
\end{array}
\right).\label{ra}\eea
%
%This coincides with (\ref{bi}), as the result of the uniform of
%(\ref{1.2}) and (\ref{3.15}).

By setting $a=\rho$\ and $b=\rho p$, where \
$p=\frac{1}{2}(-d\pm\sqrt{d^2-4})$\ with $d=\sqrt{2}$, i.e.
$p=-\exp(\pm i\pi/4)$, we regain $A$ and $B$ matrices as in
(\ref{3.4}), which satisfy the braid relation (\ref{3.5}). This
result can be also obtained through the "light-cone" limit of
(\ref{3.15}) by setting three arguments in $\breve{R}$-matrices to
be equal, i.e.
\begin{eqnarray}
u&=&v=\frac{u+v}{1+ \beta^2 uv}, \label{3.24}
\end{eqnarray}
which is satisfied by either $u=v=0$ or $u=v=\beta^{-1}$%
%\begin{eqnarray}
% \quad u=v= ^{-1}. \label{3.25}
%\end{eqnarray}
. Under the limit in (\ref{3.24}), $A(u)$\ and \ $B(u)$\ reduce to
(\ref{3.4}).
%%%page20

We know that by taking $Q\equiv i q^{-1}=i$ (i.e. $Q^4=1)$, $U$
matrix given
 by (\ref{3.22}) becomes
\bea
U(Q=i)=\frac{1}{\sqrt{2}}\left(\begin{array}{cccc}1&0&0&i\\0&1&i&0\\0&-i&1&0\\-i&0&0&1\end{array}\right),
 \label{3.26} \eea
 which is the transformation matrix for the Bell States \cite{zkg}. We thus
conclude that the 4-dimensional entangling braid matrix (\ref{3.26})
and the 2-dimensional braid matrix (\ref{3.4}) can be uniformed by
acting the TLA operator on different dimensional basis. The
4-dimensional basis can be
 $\big{(}|\upuparrows\rangle,|\uparrow\downarrow\rangle,|\downarrow\uparrow\rangle,|\downdownarrows\rangle \big{)}$,
 whereas the logic qubit basis read $\big{(}|e_1\rangle,|e_2\rangle\big{)}$. Conversely, the latter can be expanded in terms
 of four-spin states and thus relate with the 4-dimensional basis.
 In order to find this correspondence, we firstly rewrite $U_{ij}$ as a form
 of projectors
 \bea
 U_{ij}=\sqrt{2}(|\psi_{ij}\rangle\langle\psi_{ij}|+|\varphi_{ij}\rangle\langle\varphi_{ij}|),
 \eea
 where
 \bea
|\psi_{ij}\rangle=\dfrac{1}{\sqrt{2}}(|\uparrow\uparrow\rangle_{ij}+
e^{-i\phi'}|\downarrow\downarrow\rangle_{ij}),~|\varphi_{ij}\rangle=
\dfrac{1}{\sqrt{2}}(|\uparrow\downarrow\rangle_{ij}-i|\downarrow\uparrow\rangle_{ij}),
~\phi'=-(\phi+\dfrac{3\pi}{2}). \eea
It is interesting that both $|\psi_{ij}\rangle$ and
$|\varphi_{ij}\rangle$ are maximally entangled states for two spins,
i.e. Bell states. With the aid of (\ref{3.90}), we arrive at
\bea
&&|e_1\rangle=\dfrac{1}{\sqrt{1+|\nu|^2}}(|\psi_{12}\rangle|\psi_{34}\rangle+
\nu|\varphi_{12}\rangle|\varphi_{34}\rangle),\nonumber \\
&&|e_2\rangle=\dfrac{1}{\sqrt{1+|\nu|^2}}((1-i\nu
e^{i\phi'})|\psi_{23}\rangle|\psi_{41}\rangle-\nu(1-i\nu^{-1}e^{-i\phi'})
|\varphi_{23}\rangle|\varphi_{41}\rangle)-|e_1\rangle, \label{e1e2}
\eea
with $\nu$ be an arbitrary coefficient. Detailed calculation is
shown in Appendix \ref{appenb}.

Briefly, in the invariant subspace spanned by $|e_1\rangle$ and
$|e_2\rangle$, the $\breve{R}$ matrices satisfying 4D YBE
(\ref{3.15}) will reduce to 2D representation, $A(u)$ and $B(u)$,
with the corresponding 2D YBE (\ref{3.19}).
%
%\begin{figure}[ht]
%\includegraphics[width=0.35\columnwidth]{u.eps}
%\caption{By sending a photon through a QWP, then through a HWP,
%finally through another QWP, its polarization state can be changed
%unitarily to any other one. For the considered case, $A$ matrix
%corresponds to $\alpha=\eta=\dfrac{\pi}{4}$, $
% =-\dfrac{\pi}{4}+\dfrac{\theta}{2}$, and $B$ matrix corresponds
%to $\alpha=\eta=\dfrac{\pi}{2}$, $ =\dfrac{\theta}{2}$. }
%\label{wp}
%\end{figure}

\section{calculation for $|e_i\rangle$} \label{appenb}

 Here
we give the details of calculation for $|e_i\rangle(i=1,2)$. We
start from
\bea
        &&U_{12}\left|e_1\right>=U_{34}\left|e_1\right>=d\left|e_1\right>,\label{a1}\\
        &&U_{12}\left|e_2\right>=U_{34}\left|e_2\right>=0,\label{a2}\\
        &&U_{23}\left|e_1\right>=U_{14}\left|e_1\right>=\dfrac{1}{d}(\left|e_1\right>+\sqrt{d^2-1}\left|e_2\right>),\label{a3}\\
        &&U_{23}\left|e_2\right>=U_{14}\left|e_2\right>=\dfrac{\sqrt{d^2-1}}{d}(\left|e_1\right>+\sqrt{d^2-1}\left|e_2\right>).
        \label{a4}
\eea
Generally, $|e_i\rangle$ can be expanded into the linear combination
of Bell states
 \bea
|\psi_{ij}^{\pm}\rangle=\dfrac{1}{\sqrt{2}}(|\uparrow\uparrow\rangle_{ij}\pm
e^{-i\phi'}|\downarrow\downarrow\rangle_{ij}),~|\varphi_{ij}^{\pm}\rangle=\dfrac{1}{\sqrt{2}}(|\uparrow\downarrow\rangle_{ij}\mp
i|\downarrow\uparrow\rangle_{ij}). \eea
%
% Thus
%\bea
%   && |e_1\rangle = a_1 |\psi^{+}_{12}\rangle |\psi^{+}_{34}\rangle
%                 +a_2 |\psi^{+}_{12}\rangle |\psi^{-}_{34}\rangle
%                 +a_3 |\psi^{+}_{12}\rangle |\phi'^{+}_{34}\rangle
%                 +a_4 |\psi^{+}_{12}\rangle |\phi'^{-}_{34}\rangle
%                 \nonumber \\
% && ~~~~~    +a_5 |\psi^{-}_{12}\rangle |\psi^{+}_{34}\rangle
%                 +a_6 |\psi^{-}_{12}\rangle |\psi^{-}_{34}\rangle
%                 +a_7 |\psi^{-}_{12}\rangle |\phi'^{+}_{34}\rangle
%                 +a_8 |\psi^{-}_{12}\rangle |\phi'^{-}_{34}\rangle
%                     \nonumber \\
% && ~~~~~    +a_9 |\phi'^{+}_{12}\rangle |\psi^{+}_{34}\rangle
%                 +a_{10} |\phi'^{+}_{12}\rangle |\psi^{-}_{34}\rangle
%                 +a_{11} |\phi'^{+}_{12}\rangle |\phi'^{+}_{34}\rangle
%                 +a_{12} |\phi'^{+}_{12}\rangle |\phi'^{-}_{34}\rangle    \nonumber \\
% && ~~~~~    +a_{13} |\phi'^{-}_{12}\rangle |\psi^{+}_{34}\rangle
%                 +a_{14} |\phi'^{-}_{12}\rangle |\psi^{-}_{34}\rangle
%                 +a_{15} |\phi'^{-}_{12}\rangle |\phi'^{+}_{34}\rangle
%                 +a_{16} |\phi'^{-}_{12}\rangle |\phi'^{-}_{34}\rangle
%\eea
%
%and similar expansion for $|e_2\rangle$.
Due to the project form of $U_{ij}$
 \bea
 U_{ij}=\sqrt{2}(|\psi_{ij}^{+}\rangle\langle\psi_{ij}^+|+|\varphi_{ij}^+\rangle\langle\varphi^+_{ij}|),
 \eea
one can get general expression of $|e_i\rangle$ from (\ref{a1}) and
(\ref{a2})
\bea &&|e_1\rangle = a_1 |\psi^{+}_{12}\rangle |\psi^{+}_{34}\rangle
                  +a_2 |\psi^{+}_{12}\rangle |\varphi^{+}_{34}\rangle
                  +a_3 |\varphi^{+}_{12}\rangle |\psi^{+}_{34}\rangle
                  +a_{4} |\varphi^{+}_{12}\rangle
                  |\varphi^{+}_{34}\rangle, \nonumber \\
&& |e_2\rangle = a_5 |\psi^{-}_{12}\rangle |\psi^{-}_{34}\rangle
               +a_6 |\psi^{-}_{12}\rangle |\varphi^{-}_{34}\rangle
               +a_{7} |\varphi^{-}_{12}\rangle |\psi^{-}_{34}\rangle
               +a_{8} |\varphi^{-}_{12}\rangle |\varphi^{-}_{34}\rangle.
\eea
We further notice that (\ref{a3}) and (\ref{a4}) indicate the
symmetry of exchanging pair indices $23\leftrightarrow14$ for
$|e_i\rangle$. Taking of this symmetry and noticing the minus sign
in the expression of $\psi^{-}_{ij}$ and $\varphi^{+}_{ij}$, we
further simplify $|e_i\rangle$ into
\bea &&|e_1\rangle = a_1 |\psi^{+}_{12}\rangle |\psi^{+}_{34}\rangle
                   +a_{4} |\varphi^{+}_{12}\rangle
                  |\varphi^{+}_{34}\rangle, \nonumber \\
&& |e_2\rangle = a_5 |\psi^{-}_{12}\rangle |\psi^{-}_{34}\rangle
               +a_{8} |\varphi^{-}_{12}\rangle |\varphi^{-}_{34}\rangle.
\eea
Now we set $d=\sqrt{2}$ in (\ref{a3}) and (\ref{a4}), which means
$U_{23}|e_1\rangle=U_{14}|e_2\rangle$. This is the only one
condition that further determines the coefficients
$a_i~(i=1,4,5,8)$. What we proceed is detailed expansion
\bea &&U_{23}|e_1\rangle = \sqrt{2}
\{\frac{1}{2}(|\upuparrows\rangle+e^{-i\phi'}|\downdownarrows\rangle)(\langle\upuparrows|+e^{i\phi'}\langle
\downdownarrows|)+\frac{1}{2}(|\uparrow\downarrow\rangle-i|\downarrow\uparrow\rangle)(\langle\uparrow\downarrow
|+i\langle\downarrow\uparrow|)\}_{23} \nonumber \\
&&~~~~~~~~~~~~~\{
\frac{a_1}{2}(|\upuparrows\rangle+e^{-i\phi'}|\downdownarrows\rangle)_{12}(|\upuparrows\rangle+e^{-i\phi'}|
\downdownarrows\rangle)_{34}+\frac{a_4}{2}(|\uparrow\downarrow\rangle-i|\downarrow\uparrow\rangle)_{12}(|\uparrow\downarrow
\rangle-i|\downarrow\uparrow\rangle)_{34} \} \nonumber \\
&&~~~~~~~~~=|\psi_{23}^{+}\rangle \{
\frac{a_1}{2}(|\upuparrows\rangle+e^{-i\phi'}|\downdownarrows\rangle)_{14}+\frac{a_4}{2}(
-i|\downdownarrows\rangle-i e^{i\phi'}|\upuparrows\rangle)_{14} \}
\nonumber \\
&&~~~~~~~~~~~~+|\varphi_{23}^{+}\rangle \{
\frac{a_1}{2}(e^{-i\phi'}|\uparrow\downarrow\rangle+ie^{-i\phi'}|\downarrow\uparrow\rangle)_{14}+\frac{a_4}{2}(
-|\downarrow\uparrow\rangle+i|\uparrow\downarrow\rangle)_{14} \} \nonumber \\
&&~~~~~~~~~=\frac{1}{2}(a_1-ie^{i\phi'}a_4)|\psi_{23}^{+}\rangle
|\psi_{14}^+\rangle + \frac{1}{2}(a_1 e^{-i\phi'}+ia_4)
|\varphi_{23}^{+}\rangle|\varphi_{14}^-\rangle, \nonumber \\
&&U_{14}|e_2\rangle = \sqrt{2}
\{\frac{1}{2}(|\upuparrows\rangle+e^{-i\phi'}|\downdownarrows\rangle)(\langle\upuparrows|+e^{i\phi'}\langle
\downdownarrows|)+\frac{1}{2}(|\uparrow\downarrow\rangle-i|\downarrow\uparrow\rangle)(\langle\uparrow\downarrow
|+i\langle\downarrow\uparrow|)\}_{23} \nonumber \\
&&~~~~~~~~~~~~~\{
\frac{a_5}{2}(|\upuparrows\rangle-e^{-i\phi'}|\downdownarrows\rangle)_{12}(|\upuparrows\rangle-e^{-i\phi'}|
\downdownarrows\rangle)_{34}+\frac{a_8}{2}(|\uparrow\downarrow\rangle+i|\downarrow\uparrow\rangle)_{12}(|\uparrow\downarrow
\rangle+i|\downarrow\uparrow\rangle)_{34} \} \nonumber \\
&&~~~~~~~~~=|\psi_{23}^{+}\rangle \{
\frac{a_5}{2}(|\upuparrows\rangle+e^{-i\phi'}|\downdownarrows\rangle)_{14}+\frac{a_8}{2}(
i|\downdownarrows\rangle+i e^{i\phi'}|\upuparrows\rangle)_{14} \}
\nonumber \\
&&~~~~~~~~~~~~+|\varphi_{23}^{+}\rangle \{
\frac{a_5}{2}(-e^{-i\phi'}|\uparrow\downarrow\rangle-ie^{-i\phi'}|\downarrow\uparrow\rangle)_{14}+\frac{a_8}{2}(
-|\downarrow\uparrow\rangle+i|\uparrow\downarrow\rangle)_{14} \} \nonumber \\
&&~~~~~~~~~=\frac{1}{2}(a_5+ie^{i\phi'}a_8)|\psi_{23}^{+}\rangle
|\psi_{14}^+\rangle + \frac{1}{2}(-a_5 e^{-i\phi'}+ia_8)
|\varphi_{23}^{+}\rangle|\varphi_{14}^-\rangle. \eea
Because of the orthogonality of Bell states, we get the relation
between coefficients
\bea && a_1 - i e^{i\phi'} a_4 =a_5+ i e^{i \phi'} a_8, \nonumber \\
 &&      a_1 e^{-i\phi'}+ia_4 = -a_5 e^{-i\phi'}+ia_8, \nonumber \\
 && \Rightarrow ~ a_5=-ie^{i\phi'}a_4,~a_8=-ie^{-i\phi'}a_1.
 \eea
Setting $a_1=\dfrac{1}{\sqrt{1+|\nu|^2}}$ and
$a_4=\dfrac{\nu}{\sqrt{1+|\nu|^2}}$, we finally arrive at
\bea
&&|e_1\rangle=\dfrac{1}{\sqrt{1+|\nu|^2}}(|\psi_{12}^+\rangle|\psi_{34}^+\rangle+
\nu|\varphi_{12}^+\rangle|\varphi_{34}^+\rangle),\nonumber \\
&&|e_2\rangle=\dfrac{-i}{\sqrt{1+|\nu|^2}}(\nu
e^{i\phi'}|\psi_{12}^-\rangle|\psi_{34}^-\rangle+
e^{-i\phi'}|\varphi_{12}^-\rangle|\varphi_{34}^-\rangle). \eea
They are in deed equivalent to (\ref{e1e2}). The arbitrary parameter
$\nu$ represents a certain degeneracy between the components of
$|e_i\rangle$ with respect to the actions of $U_{ij}$. From the
process of calculation, we can view $U_{23}$ and $U_{14}$ as the
entanglement swapping operators on the pair-entangled states
$|e_i\rangle$, in accord with the results in \cite{ckg}.

\section{Decomposition of $\breve{R}(\theta)$} \label{appenc}

Here we give the proof of the decomposition (\ref{dec}) based on the
work of S. S. Bullock and his coauthors \cite{dec,bullock}. They
have developed the following criterion

\emph{Proposition 1.} An operator $u\in SU(4)$ can be simulated
using no CNOT gates and arbitrary one-qubit gates from SU(2) iff
$\chi[\gamma(u)]=(x\pm1)^4$. Here $\gamma(u)=u(\sigma_y
\otimes\sigma_y)u^{T}(\sigma_y \otimes\sigma_y)$, $u^T$ denotes the
transpose and $\chi[g]=\mathrm{det}[xI-g]$ denotes the
characteristic polynomial of $g$.

\emph{Proposition 2.} An operator $u\in SU(4)$ can be simulated
using one CNOT gates and arbitrary one-qubit gates from SU(2) iff
$\chi[\gamma(u)]=(x+i)^2(x-i)^2$.

\emph{Proposition 3.} An operator $u\in SU(4)$ can be simulated
using two CNOT gates and arbitrary one-qubit gates from SU(2) iff
$\mathrm{tr}[\gamma(u)]=$ is real.

Direct calculation shows that the case $\theta=0$ satisfies
Proposition 1 while the case $\theta=\pi/4$ or $3\pi/4$ satisfies
Proposition 2. Since
$\chi[\gamma(\breve{R}(\theta))]=\big{(}1+x^2-2x\cos2\theta\big{)}$
and $\mathrm{tr}[\gamma(\breve{R}(\theta))]=4\cos 2\theta$, so
Proposition 3 confirms that $\breve{R}(\theta)$ generally admits a
quantum circuit using two CNOT gates. The explicit form of $V_i$ in
(\ref{ui}) is calculated by the algorism in \cite{dec}.

\begin{figure}[ht]
\includegraphics[width=0.8\columnwidth]{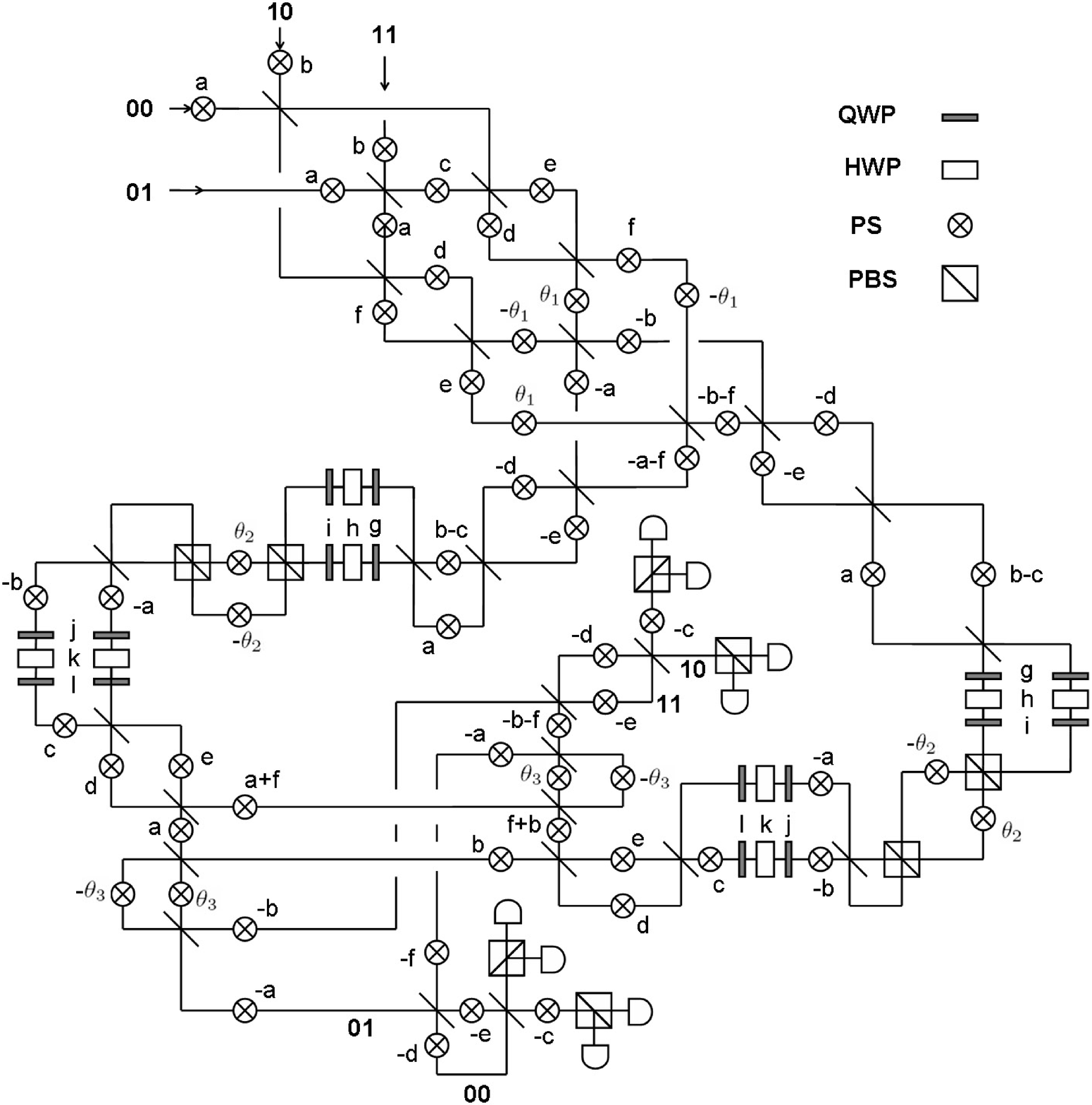}
\caption{The whole optical setup of simulating LHS of 4D YBE shown
in Fig.\ref{ybe4}. The binary numbers indicate location qubit basis.
Each pair of $-\pi/2$ phase shifters accompanied with every beam
splitter is not shown. The phase shifts of other PSs, from a to f,
are $-(\pi+\phi)/4$, $(5\pi+\phi)/4$, $\pi/2$, $(\phi-\pi)/4$,
$(\phi-3\pi)/4$ and $(\pi-\phi)/2$, respectively. The angles of wave
plates to form $U_2$ and $U_2^{\dagger}$, from g to l, are $\pi/2$,
$(\pi-\phi)/8$, $-\pi/4$, $\pi/4$, $(5\pi-\phi)/8$ and $0$,
respectively (see (\ref{uii})). Mirrors are placed on every corner.}
\label{total}
\end{figure}

\end{document}